\documentclass{iau}
\usepackage{graphicx}

\title[] 
{Terrestrial planet formation in low-mass disks: dependence with initial conditions}

\author[Ronco, de El\'{\i}a, Guilera] 
{M. P. Ronco$^{1}$, G. C. de El\'{\i}a$^{1}$ \& O. M. Guilera$^{1}$}

\affiliation{$^1$Grupo de Ciencias Planetarias, Facultad de Ciencias Astron\'omicas y Geof\'{\i}sicas \& Instituto de Astrof\'{\i}sica de La Plata (CONICET - UNLP), Argentina. \\ email: {\tt mpronco@fcaglp.unlp.edu.ar}\\[\affilskip]}

\pubyear{2014}
\volume{310}  
\pagerange{xxx--xxx}
\jname{Complex Planetary Systems}
\editors{Z. Knezevic \& A. Lema\^itre}
\begin{document}

\maketitle

\begin{abstract}
In general, most of the studies of terrestrial-type planet formation typically use ad hoc initial conditions. In this 
work we improved the initial conditions described in \cite[Ronco \& de El\'{\i}a (2014)]{Ronco \& de El\'{\i}a2014} 
starting with a semi-analytical model wich simulates the evolution of the protoplanetary disk during the gas phase. 
The results of the semi-analytical model are then used as initial conditions for the N-body simulations. We show that the 
planetary systems considered are not sensitive to the particular initial distribution of embryos and planetesimals and thus, 
the results are globally similar to those found in the previous work.
 
\end{abstract}

\firstsection 
\section{Introduction}

Many observational and theoretical works suggest that planetary systems with only rocky 
planets are the most common in the Universe. In particular, \cite[Miguel et al. (2011)]{Miguel.et.al.2011} indicated that 
a planetary system with only small rocky planets is the most common outcome obtained from a low-mass disk 
($\lesssim$ 0.03 $M_{\odot}$) for different surface density profiles. In general, most of the studies of terrestrial-type planet 
formation typically use ad hoc initial conditions (\cite[Kokubo \& Ida, 1998]{Kokubo.et.al.1998}; \cite[Raymond et al. 2005]{Raymond.et.al.2005}). Here, 
we complement the results of 
\emph{N-body high-resolution simulations} performed by \cite[Ronco \& de El\'{\i}a (2014)]{Ronco \& de El\'{\i}a2014} 
starting from a semi-analytical model developed by \cite[Brunini \& Benvenuto (2008)]{Brunini.and.Benvenuto2008} and 
\cite[Guilera et al. (2010)]{Guilera.et.al.2010} wich simulates the evolution of the protoplanetary disk during the gas phase. 
We analyze the formation of terrestrial planets and water delivery without gas giants and compare
the results with the previous work.

The semi-analytical model is used to calculate the formation of several embryos between 0.5~AU and 5~AU (as in \cite[de El\'{\i}a et al. 2013]{deElia.et.al.2013}). These embryos were 
separated by 10 mutual Hill radii and their initial masses correspond to the transition mass between runaway and oligarchic 
growth (\cite[Ida \& Makino, 1993]{Ida.and.Makino1993}). In our previous work we adopted three different values for the exponent $\gamma$ that characterize the slope of the surface density ($\gamma = 0.5$, $1$ and $1.5$). The planetary systems formed 
with $\gamma = 1.5$ were the most distinctive ones from an astrobiological point of view. Thus, in this new work we developed three 
N-body simulations with new initial conditions given by the semi-analytical model, only for $\gamma = 1.5$. For this 
profile we found the same proportion for both populations: half the mass in embryos and half the mass in planetesimals after the gas 
is completely dissipated in 3~Myr.

\section{Results}

\begin{figure}[t]
\vspace*{-0.5 cm}
\begin{center}
 \includegraphics[width=2.03in]{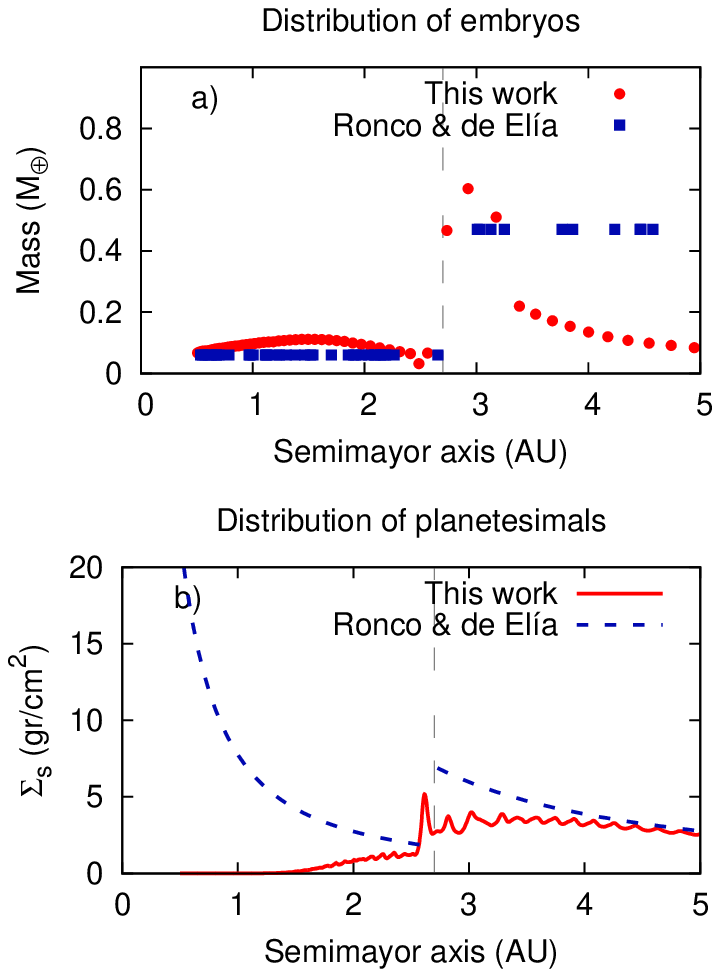} 
\includegraphics[width=2.8in]{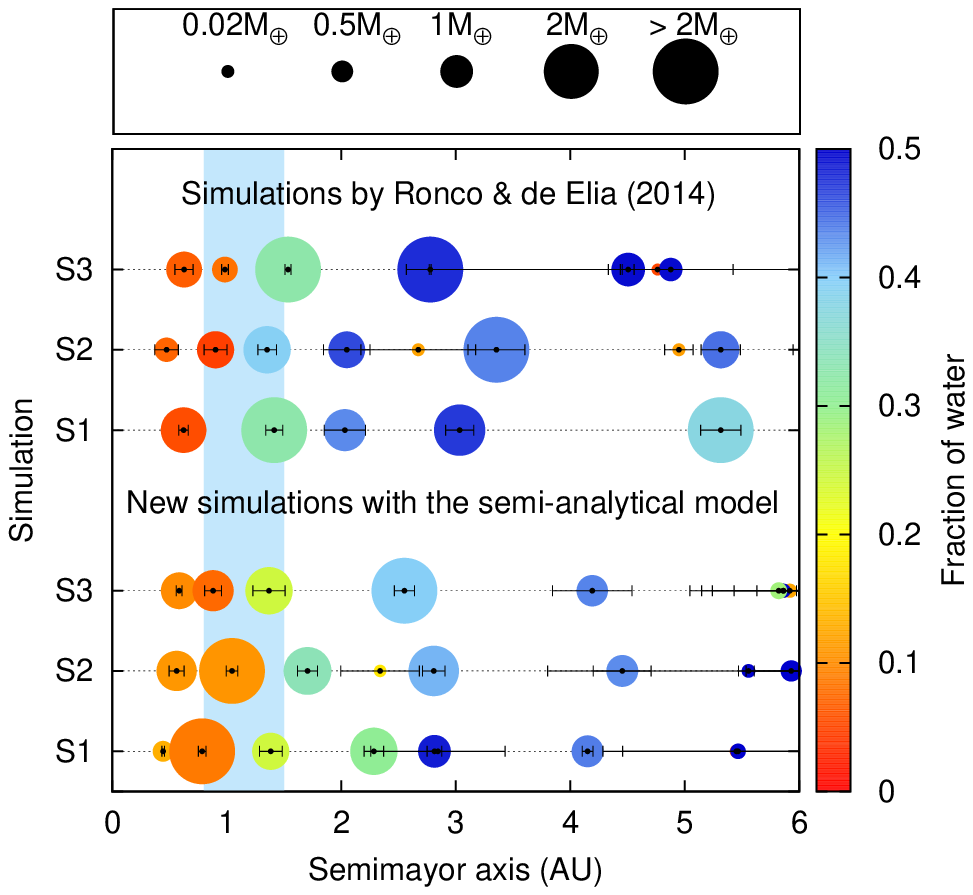} 
 \vspace*{-0.1 cm}
 \caption{Left: a) Distributions of embryos used to start the N-body simulations. The squares represent the distribution 
of embryos used by \cite[Ronco \& de El\'{\i}a (2014)]{Ronco \& de El\'{\i}a2014} and the circles represent the final results obtained 
with the semi-analytical model. b) Surface density profiles used to distribute 1000 planetesimals to start the N-body simulations. 
The dashed line represents the surface density used in \cite[Ronco \& de El\'{\i}a (2014)]{Ronco \& de El\'{\i}a2014} and the solid line 
represents the final results obtained with the semi-analytical model. Right: Final configuration of the simulations obtained in 
\cite[Ronco \& de El\'{\i}a (2014)]{Ronco \& de El\'{\i}a2014} and the new ones obtained with the 
semi-analytical model. The color scale represent the water content and the shaded region, the HZ. The excentricity of 
each planet is shown over it, by its radial movement over an orbit. Color figure only available in the electronic version.}
 \label{fig1}
\end{center}
\vspace*{-0.1 cm}
\end{figure}

These new planetary systems are globally similar to those found in our previous work. Each simulation formed 
between 1 and 2 planets in the habitable zone (HZ) (between 0.8~AU and 1.5~AU) after 200~Myr of evolution (Fig. \ref{fig1}). Their masses 
range between $1.18M_\oplus$ and $2M_\oplus$ and their water content between 7.5\% and 24.3\% by mass, which represent between 427 and 1671 
Earth oceans (1 Earth ocean $=$ $2.8\times 10^{-4}M_\oplus$). We also found planets with masses from $1M_\oplus$ to $2.36M_\oplus$ 
near the snow line (located at 2.7~AU), which can be discovered by the microlensing technique. The masses of the planets in the HZ 
are large enough to retain an atmosphere and to sustain plate tectonics, and as we also formed in the previous simulations, this profile 
formed \emph{water worlds} that come from beyond the snow line. Thus, the planets that remain in the HZ present the characteristics 
to be potencially habitable.

We therefore conclude that the results are globally similar to those found by \cite[Ronco \& de El\'{\i}a (2014)]{Ronco \& de El\'{\i}a2014}. These planetary systems do not seem to be sensitive to the particular initial distribution of embryos and planetesimals and we suggest that the strong dependence on the final results would go with the initial mass proportion used in both populations. However, these more realistic initial conditions allowed us to find more reliable results concerning the water delivery and the global dinamics of the planetary systems.

\end{document}